\documentclass[aps,prb,twocolumn,showpacs]{revtex4}
\usepackage[dvips]{graphicx}

\begin{document}

\title{Spin dynamics in the antiferromagnetic phase of electron-doped
cuprate superconductors}
\author{Qingshan Yuan,$^{1,2}$ T. K. Lee,$^3$ and C. S. Ting$^1$}
\address{$^1$Texas Center for Superconductivity and Advanced Materials 
and Department of Physics, University of Houston, Houston, TX 77204\\
$^2$Pohl Institute of Solid State Physics, Tongji University,
Shanghai 200092, P. R. China\\
$^3$Institute of Physics, Academia Sinica, Nankang, Taipei, Taiwan 11529
}

\begin{abstract}
Based on the $t$-$t'$-$t''$-$J$ model we have calculated the dynamical
spin susceptibilities in the antiferromagnetic (AF) phase for
electron-doped cuprates, by use of the slave-boson mean-field theory
and random phase approximation.
Various results for the susceptibilities versus energy and momentum
have been shown at different dopings. At low energy, except the collective 
spin-wave mode around $(\pi,\pi)$ and $0$, we have primarily observed that
resonance peaks will appear around $(0.3\pi,0.7\pi)$ and equivalent points
with increasing doping, which are due to the single particle-hole excitations
between the two AF bands. The peaks are pronounced in the transverse susceptibility
but not in the longitudinal one. These features are predicted for neutron
scattering measurements.
\end{abstract}


\maketitle

Since their discovery the hole- and electron-doped high-$T_c$ cuprate
superconductors have been noticed to exhibit many different properties.
This is immediately seen from the phase diagram in the 
temperature/doping plane. In hole-doped compounds, e.g.,
La$_{2-x}$Sr$_x$CuO$_4$, the antiferromagnetic (AF) order stabilizes
only in a narrow doping region $x\le 0.02$,\cite{Takagi} whereas it persists
up to $x=0.14$ in electron-doped ones, e.g.,
Nd$_{2-x}$Ce$_x$CuO$_4$.\cite{Luke}
The electron-hole asymmetry has also been observed in various
measurements such as nuclear magnetic resonance,\cite{Zheng}
inelastic neutron scattering,\cite{Yamada} etc. Particularly,
recent angle-resolved photoemission spectroscopy (ARPES) experiments
\cite{Armitage,Damascelli} have revealed the peculiar Fermi surface (FS) 
structure in electron-doped cuprate Nd$_{2-x}$Ce$_x$CuO$_4$. 
It was found that at low doping the FS is a small pocket centered at $(\pi,0)$,
in contrast to the hole-doped case\cite{Yoshida} where it is around 
$(\pi/2,\pi/2)$. Moreover, upon increased doping 
another pocket begins to form around
$(\pi/2,\pi/2)$ and eventually at optimal doping $x=0.15$ 
the several FS pieces together constitute a large curve around $(\pi,\pi)$.

Theoretically, the essential role of the next nearest neighbor (n.n.)
hopping $t'$ has been pointed out to understand the electron-hole
asymmetry.\cite{Tohyama94,Gooding}
By inclusion of $t'$ and possibly the third n.n. hopping $t''$ in two types
of models $t$-$J$ and $t$-$U$, numerous numerical and analytical studies
have been carried out to explore the different properties of electron-doped
cuprates.\cite{Tohyama01,Lee,Li,Tohyama04,Kusko,Kusunose,Kyung,Yuan}
Much attention has been paid to the interpretation\cite{Kusko,Kusunose,Kyung,Yuan}
of the FS evolution with doping as observed by ARPES.
\cite{Armitage,Damascelli}
By use of the $t$-$t'$-$t''$-$U$ model Kusko {\it et al.}\cite{Kusko}
have derived the mean-field (MF) quasiparticle energy bands in the AF state. 
In order to get the consistent results with ARPES, however,
they need to introduce a doping-dependent effective parameter $U$, 
see also Ref.~[\onlinecite{Kyung}].
Alternatively, we have adopted the $t$-$t'$-$t''$-$J$ model to construct
the FS.\cite{Yuan} The use of $t$-$J$-type models for the electron-doped 
cuprates is a natural generalization from their extensive application to
the hole-doped ones, and is largely stimulated by the accumulating evidence 
for the universal $d$-wave superconducting (SC) gap\cite{Tsuei}
in both kinds of materials.
Without phenomenological parameters as argued by Kusko {\it et al.},
we have obtained that at low doping only one AF band is crossed by
the Fermi level around $(\pi,0)$, and upon increased doping 
the other one will again be crossed around $(\pi/2,\pi/2)$.
Correspondingly a FS pocket forms initially around $(\pi,0)$ and later
the new one appears around $(\pi/2,\pi/2)$, in agreement with the ARPES data.

In view of the sample quality, on the other hand, it may be questioned
whether the multiple pockets around the inequivalent points revealed by ARPES 
are detected from the uniform phase of the whole sample
or from the different regions of the inhomogenous sample.
Thus other experimental measurements complementary to ARPES are strongly needed.
Naturally, any theoretical prediction based on the current result
will be useful to guide the experiments.
For this purpose the spin dynamics, which can be measured by
inelastic neutron scattering, is calculated in this paper.
If it is really true that the new FS emerges in the same uniform phase
with the old one upon increased doping, the novel particle-hole ({\it p-h})
excitations between them should lead to characteristic features for the
spin dynamics which implies the information of the spin/charge excitations.

At present, the calculations on the dynamical spin susceptibilities
for electron-doped cuprates\cite{Li,Tohyama04,Onufrieva} are
very limited, and mostly done in the SC and normal states.
Here we will concentrate on the AF phase, and
study the variation of the susceptibilities
with doping which has not yet been investigated.
The significance of the topic is highlighted in the electron-doped materials
because the AF phase is robust to survive a wide doping range
and the understanding of its nature becomes crucial.
Thus our motivation is twofold.
On one hand, we wish to predict some characteristic results for experimental
verification. On the other hand, we perform the general formulation of
the dynamical spin susceptibilities under the background of the AF order,
which is rarely presented in the literature.\cite{Schrieffer} 
The random phase approximation (RPA) is
used to take the spin fluctuation into account.
Our main result is that with increasing doping
new resonance peaks will appear at low energy around
$(0.3\pi,0.7\pi)$ and equivalent points, which are pronounced
in the transverse susceptibility but not in the longitudinal one.

We begin with the $t$-$t'$-$t''$-$J$ model Hamiltonian
\begin{eqnarray}
H & = & -t\sum_{\langle ij\rangle \sigma}(c_{i\sigma}^{\dagger}c_{j\sigma}+{\rm h.c.})
-t'\sum_{\langle ij\rangle_2\sigma}(c_{i\sigma}^{\dagger}c_{j\sigma}+{\rm h.c.})
\nonumber\\
& & -t''\sum_{\langle ij\rangle_3 \sigma}(c_{i\sigma}^{\dagger}c_{j\sigma}+{\rm h.c.})
+J\sum_{\langle  ij\rangle }({\bf S}_i \cdot {\bf S}_j-{1\over 4}n_i n_j)\nonumber\\
& & -\mu_0 \sum_{i\sigma}c_{i\sigma}^{\dagger}c_{i\sigma}\ ,\label{H}
\end{eqnarray}
where all the notation is standard.
For electron-doping, one has 
$t<0,\ t'>0$ and $t''<0$.\cite{Tohyama94,Gooding,Tohyama01,Lee}
Throughout the work $|t|$ is taken as the energy unit.
Typical values are adopted: $t'=0.3,\ t''=-0.2$ and $J=0.3$.

We treat the Hamiltonian (\ref{H}) by the slave-boson MF theory. Without details
we briefly introduce the necessary formulas.
Under assumption of boson condensation and definition of MF parameters:
the staggered magnetization $m=(-1)^i\langle S_i^z\rangle$ and
the uniform bond order $\chi=\langle  f_{i\sigma}^{\dagger}f_{j\sigma}\rangle$
($f$: spinon operator),
the Hamiltonian (\ref{H}) is decoupled as follows in momentum space\cite{Yuan}
\begin{eqnarray}
H_{M} & = & {\sum_{k,\sigma}}' (\varepsilon_k f_{k\sigma}^{\dagger}f_{k\sigma}
+ \varepsilon_{k+Q} f_{k+Q\sigma}^{\dagger}f_{k+Q\sigma})
-2Jm\ \nonumber\\
& & {\sum_{k,\sigma}}' \sigma (f_{k\sigma}^{\dagger}f_{k+Q\sigma}
+{\rm h.c.})+2NJ(\chi^2+m^2)\ ,\label{Hk}
\end{eqnarray}
where $\varepsilon_k=(2|t|x-J\chi) (\cos k_x+\cos k_y)
-4t'x\cos k_x \cos k_y-2t''x(\cos 2k_x+\cos 2k_y)-\mu$
($x$: doping concentration), $Q=(\pi,\pi)$, and $N$ is the total
number of lattice sites. $\sum'_k$ means the summation over only the
magnetic Brillouin zone (MBZ): $-\pi<k_x\pm k_y\le \pi$.
Above, the local constraint for no double occupancy has been treated
in average as usually done. In addition, we do not consider the
potential pairing in order to harmonize with the experimental observation
for the large range of pure AF phase.

By use of the unitary transformations:
$f_{k\sigma}=\cos \theta_k \alpha_{k\sigma}+\sigma\sin \theta_k\beta_{k\sigma}$
and
$f_{k+Q\sigma}=-\sigma\sin \theta_k \alpha_{k\sigma}+
\cos \theta_k\beta_{k\sigma}$ [$\sigma=\uparrow(+)$ or $\downarrow(-)$],
where
$\cos 2\theta_k = (\varepsilon_{k+Q}-\varepsilon_{k})/\gamma_k$,
$\sin 2\theta_k = -4Jm/\gamma_k$ and
$\gamma_k =\sqrt{(\varepsilon_{k+Q}-\varepsilon_{k})^2+(4Jm)^2}$,
the Hamiltonian (\ref{Hk}) can be diagonalized in terms of $\alpha_{k\sigma}$
and $\beta_{k\sigma}$, with the two energy bands
$\xi_{k,\alpha} = (\varepsilon_{k}+\varepsilon_{k+Q}-\gamma_k)/2$ and
$\xi_{k,\beta}  = (\varepsilon_{k}+\varepsilon_{k+Q}+\gamma_k)/2$.

For each given doping $x$ and temperature $T$, the MF parameters $m$ and $\chi$,
as well as the chemical potential $\mu$ are calculated self-consistently.
Then the energy bands and the corresponding FS can be plotted in the MBZ.
All the results have been shown in Ref.~[\onlinecite{Yuan}] for $T=10^{-3}$.
Mainly, with increasing doping the two AF bands become close to
each other due to the decreasing magnetization. Within a very narrow doping range
around $x=0.144$ they are both crossed by the Fermi level around $(\pi,0)$
(and equivalent points) and $(\pm\pi/2,\pm\pi/2)$, respectively,
leading to multiple FS pockets around the inequivalent points.
In order to reveal their consequence to the observable physical
quantities, we calculate the spin dynamics in the following.

The spin susceptibilities (transverse and longitudinal) are defined by
\begin{eqnarray}
\chi_{(0)}^{+-(zz)}(q,q',\tau) & = & +{1\over N} \langle T_{\tau} S_q^{+(z)}(\tau) 
S_{-q'}^{-(z)}(0)\rangle_{(0)}\ ,
\end{eqnarray}
where $\langle\cdots\rangle_{(0)}$ means thermal average on the eigenstates
of $H_{(M)}$, $S_q^+ =\sum_i S_i^+ e^{iq\cdot R_i}=
\sum_k f_{k+q\uparrow}^{\dagger}f_{k\downarrow}$\ , $S_q^- =(S_{-q}^+)^{\dagger}$
and $S_q^z ={1\over 2}\sum_{k\sigma} \sigma f_{k+q\sigma}^{\dagger}f_{k\sigma}$.
Correspondingly the term $J\sum_{\langle  ij\rangle }{\bf S}_i \cdot {\bf S}_j$
is rewritten as
$(1/N)\sum_{q} J(q) (S_q^+ S_{-q}^- + S_q^z S_{-q}^z)$
with $J(q)=J(\cos q_x + \cos q_y)$.

We first calculate $\chi_0^{+-(zz)}$ under the quadratic Hamiltonian $H_{M}$.
By transforming $f$ operator into $\alpha$ and $\beta$ ones
we finally obtain
\begin{eqnarray}
\chi_0^{+-}(q,q',i\omega_n) & = & \delta_{q',q}\chi_0^{+-}(q,i\omega_n)+\nonumber\\
 & & \delta_{q',q+Q}\chi_0^{+-}(q,q+Q,i\omega_n)\ ,\label{chi0+-qq'}\\ 
\chi_0^{zz}(q,q',i\omega_n) & = & \delta_{q',q}\chi_0^{zz}(q,i\omega_n)\ ,
\label{chi0zzqq'}
\end{eqnarray}
where
\begin{widetext}
\begin{eqnarray}
\chi_0^{+-}(q,i\omega_n) & = & -{1\over N}{\sum_k}'
\left[ \cos^2 (\theta_k+\theta_{k+q}) (F_{\alpha\alpha}+F_{\beta\beta})
+ \sin^2 (\theta_k+\theta_{k+q}) (F_{\alpha\beta}+F_{\beta\alpha}) \right]\ ,
\label{chi0+-}\\
\chi_0^{+-}(q,q+Q,i\omega_n) & = & {1\over 2N}{\sum_k}'
\left[ (\sin 2\theta_{k+q}-\sin 2\theta_k) 
(F_{\alpha\alpha}-F_{\beta\beta})
+(\sin 2\theta_{k+q}+\sin 2\theta_k)
(F_{\alpha\beta}-F_{\beta\alpha}) \right]\ ,\label{chi0Q}\\
\chi_0^{zz}(q,i\omega_n) & = & -{1\over 2N}{\sum_k}'
\left[ \cos^2 (\theta_k-\theta_{k+q}) (F_{\alpha\alpha}+F_{\beta\beta})
+ \sin^2 (\theta_k-\theta_{k+q}) (F_{\alpha\beta}+F_{\beta\alpha}) \right]
\label{chi0zz}
\end{eqnarray}
\end{widetext}
with abbreviations
$$ F_{\eta\eta'} = {n(\xi_{k+q,\eta})-n(\xi_{k,\eta'})\over 
i\omega_n+\xi_{k+q,\eta}-\xi_{k,\eta'}}\ \ (\eta,\eta'=\alpha,\beta)\ .
$$
Above $\omega_n$ are Matsubara frequencies and $n(\cdots)$ is the Fermi function.
Since the eigenstates of $H_{M}$ exhibit the AF order that breaks the spin rotational
symmetry, the simple relation $\chi_0^{+-}=2\chi_0^{zz}$ which holds 
in the case of $m=0$ is no longer valid. Moreover, 
the nondiagonal correlation function $\chi_0^{+-}$ with $q'=q+Q$ [Eq.~(\ref{chi0Q})] 
arises due to the umklapp processes.
It is instructive to have a look at the physics implied by the diagonal functions
$\chi_0^{+-(zz)}(q,i\omega_n)$ from the above Eqs.~(\ref{chi0+-}) and (\ref{chi0zz}).
With momentum transfer $q$ there are two types of {\it p-h} excitations,
either within the single band ($\alpha$ or $\beta$) as described
by $F_{\alpha\alpha}$ and $F_{\beta\beta}$ or between the two bands
by $F_{\alpha\beta}$ and $F_{\beta\alpha}$.
We notice that the two, intraband and interband {\it p-h} excitations,
have different weights in their contribution to $\chi_0^{+-(zz)}(q,i\omega_n)$.
Also, the difference between $\chi_0^{+-}(q,i\omega_n)$ and $\chi_0^{zz}(q,i\omega_n)$
exists in their different weighing factors for each type of excitations.
If we consider low temperature ($T\rightarrow 0$) and 
low energy transfer ($\omega\rightarrow 0$), the dominant contribution
to $\chi_0^{+-(zz)}(q,i\omega_n)$ comes from those excitations within the very
vicinity of the FS. Thus we expect that new resonance peaks with large momentum
transfer will arise with increasing doping due to the emergence of the
new FS pockets around $(\pm\pi/2,\pm\pi/2)$.

We further calculate $\chi^{+-(zz)}(q,q',i\omega_n)$. The residual
fluctuation of the $J$-term beyond MF is considered by the RPA.
We have obtained the similar equations for $\chi^{+-(zz)}(q,q',i\omega_n)$ 
to Eqs.~(\ref{chi0+-qq'}) and (\ref{chi0zzqq'}) for $\chi_0^{+-(zz)}(q,q',i\omega_n)$,
with correspondingly
\begin{widetext}
\begin{eqnarray}
\chi^{+-}(q,i\omega_n) = &\nonumber\\
\frac{\chi_0^{+-}(q,i\omega_n)
-J(q+Q)[\chi_0^{+-}(q+Q,q,i\omega_n)\chi_0^{+-}(q,q+Q,i\omega_n)-
\chi_0^{+-}(q,i\omega_n)\chi_0^{+-}(q+Q,i\omega_n)]}
{[1+J(q)\chi_0^{+-}(q,i\omega_n)][1+J(q+Q)\chi_0^{+-}(q+Q,i\omega_n)]-
J(q)J(q+Q)\chi_0^{+-}(q+Q,q,i\omega_n)\chi_0^{+-}(q,q+Q,i\omega_n)} &\ ,
\label{chi}
\end{eqnarray}
\end{widetext}
\begin{eqnarray}
\chi^{+-}(q,q+Q,i\omega_n) & = &
\frac{\chi_0^{+-}(q,q+Q,i\omega_n)}{{\rm denominator\ of\ Eq.~(\ref{chi})}}\ ,\\
\chi^{zz}(q,i\omega_n) & = & {\chi_0^{zz}(q,i\omega_n)\over 
1+J(q)\chi_0^{zz}(q,i\omega_n)}\ .
\end{eqnarray}

\begin{figure}[ht]
\begin{center}
\includegraphics[width=8.6cm,height=16cm,clip]{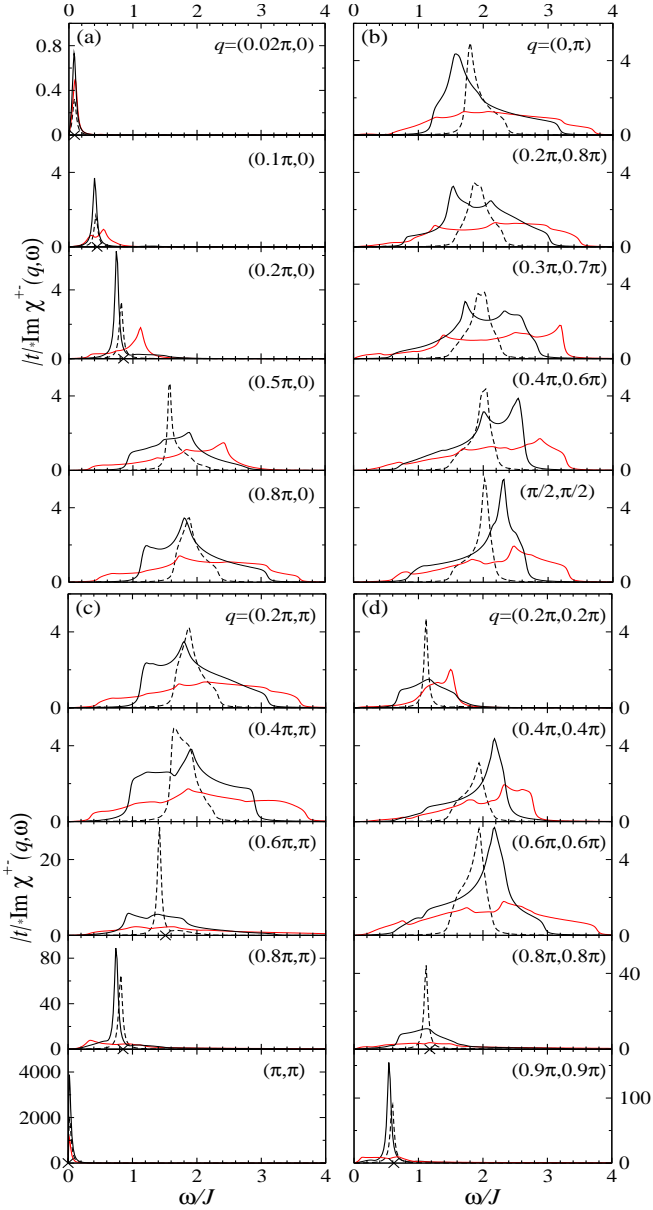}
\end{center}
\vspace*{-3mm}
\caption{The transverse dynamical susceptibility vs frequency
for various wave vectors: (a) $q_y=0$,
(b) $q_x+q_y=\pi$, (c) $q_y=\pi$, and (d) $q_x=q_y$. The dashed, solid, and
red lines in each panel represent the results for doping $\delta=0.04$, $0.1$,
and $0.144$, respectively. Note that all the values for $\delta=0.04$
have been reduced by three times.
The crosses shown in part of the panels $q\rightarrow 0$ and $Q$
denote the single magnon excitation energies
for the Heisenberg model: $J\sqrt{4-(\cos q_x +\cos q_y)^2}$. 
The parameters are: temperature $T=10^{-3}$ and damping rate 
$\Gamma=10^{-2}$ (in units of $|t|$).}
\label{Fig:CHpw}
\end{figure}

\begin{figure}[ht]
\begin{center}
\includegraphics[width=8.8cm,height=4.2cm,clip]{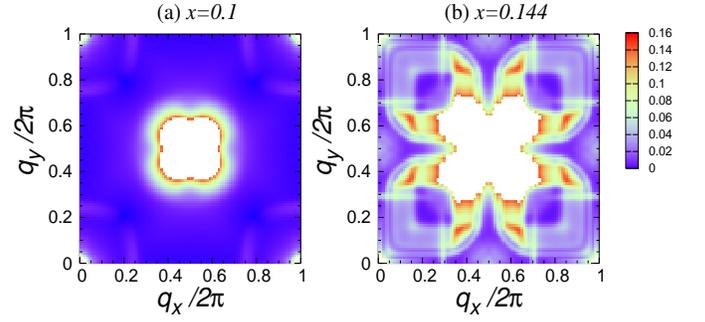}
\end{center}
\vspace*{-3mm}
\caption{The density plots of Im$\chi^{+-}(q,\omega)$ in the $q$ plane
for doping $x=0.1$ and $0.144$, with fixed low energy $\omega=0.1J$.
The white region in each panel contains a huge peak
around $(\pi,\pi)$ originating from the collective spin-wave excitation,
which is not plotted in order to highlight the much weaker peaks
at other wave vectors. With increasing doping, new peaks around
$q/2\pi\simeq (0.15,0.37)$ and equivalent points are clearly seen.}
\label{Fig:CHpq}
\end{figure}

\begin{figure}[ht]
\begin{center}
\includegraphics[width=8.6cm,height=9cm,clip]{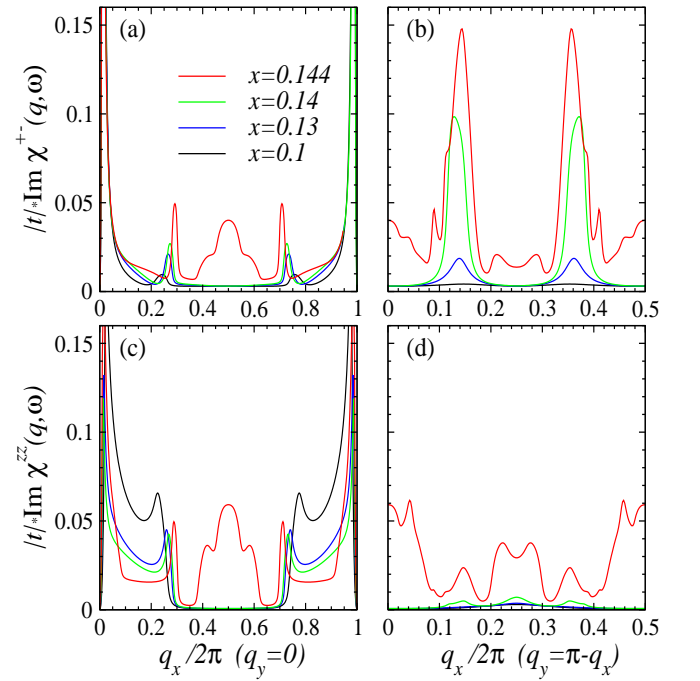}
\end{center}
\vspace*{-3mm}
\caption{Im$\chi^{+-}(q,\omega)$ [(a) and (b)] and
Im$\chi^{zz}(q,\omega)$ [(c) and (d)] at $\omega=0.1J$
with change of $q$ along the lines: (i) $q_y=0$ [(a) and (c)], 
(ii) $q_x+q_y=\pi$ [(b) and (d)].}
\label{Fig:CHp_w003}
\end{figure}

The formula for $\chi^{+-}(q,i\omega_n)$ becomes complicated due to
the existence of the nondiagonal $\chi_0^{+-}$. As a result, it may contain
a pole, which corresponds to a collective spin-wave excitation mode.
The imaginary part of $\chi^{+-}(q,i\omega_n)|_
{i\omega_n\rightarrow \omega+i\Gamma}$ is shown by Fig.~\ref{Fig:CHpw}
as a function of $\omega$ for various $q$ vectors.
Throughout the calculation we have taken
$T=10^{-3}|t|$, the damping rate $\Gamma=0.01|t|$, and 
$500\times 500$ $k$ points in the MBZ.
Three typical doping values are adopted in Fig.~\ref{Fig:CHpw}:
small $x=0.04$, medium $x=0.1$ and 
$x=0.144$ where new FS pockets around $(\pm\pi/2,\pm\pi/2)$ appear.
Globally it is seen that with increasing doping
the visible susceptibility spans a wider energy range,
due to the broadening bandwidth and reducing AF gap.
For details, we first look at the region $q\rightarrow Q=(\pi,\pi)$ shown by the 
several lower panels of Figs.~\ref{Fig:CHpw}(c) and \ref{Fig:CHpw}(d).
It is observed that a sharp peak, particularly for doping $x=0.04$ (note 
three times reduction of the amplitude in this case),
is formed and becomes stronger when $q$ is closer to $Q$.
The peak position in each panel is around the magnon excitation energy
for the corresponding Heisenberg model as shown by the cross on the abscissa.
This indicates the collective spin-wave excitation in the presence of carriers.
The peak is more prominent for $x=0.04$ because the collective mode is better defined
for smaller doping. Similarly, the peak from the spin-wave excitation is present
at low energy when $q\rightarrow 0$, as seen from the several upper panels of
Fig.~\ref{Fig:CHpw}(a).

Then we come to $q$ away from $Q$ and $0$. Now the collective
excitation has a sizable energy. We notice that this mode may not always be 
exhibited by the RPA, for example, in the line: $q_x+q_y=\pi$ one has
$\chi^{+-}=\chi_0^{+-}$ because of $J(q)=0$.
However, our interest is the experimentally relevant
low-energy region where the single {\it p-h} excitations play the essential role
when $q$ is away from $Q$ and $0$.
A notable feature observed at low $\omega$ ($<0.4J\sim 40$ meV) is that
the susceptibility becomes finite when $x=0.144$
for some wave vectors, e.g., $q=(0.3\pi,0.7\pi)$ as shown in
Fig.~\ref{Fig:CHpw}(b). To fully view the susceptibility
in the whole $q$ plane, we have fixed the low energy $\omega=0.1J$ and
plotted the density of Im$\chi^{+-}(q,\omega)$ in Fig.~\ref{Fig:CHpq}.
There is a huge peak around $Q$ in each panel which has been left blank.
With increasing doping, except an expansion of the blank region indicating
the broadening of the peak around $Q$, we can observe the considerable
intensity around the new wave vectors $q/2\pi\simeq (0.15,0.37)$ and
equivalent points [keep in mind the symmetry of $\chi^{+-(zz)}$ with respect to
$q$: $q_x\leftrightarrow q_y$ or $q_{x(y)}\rightarrow -q_{x(y)}+2\pi$].
To more clearly see the appearance of the new peaks, we have calculated 
Im$\chi^{+-}(q,\omega)$
vs $q$ along a few lines in the $q$ plane for different dopings,
which are shown by Figs.~\ref{Fig:CHp_w003}(a) and \ref{Fig:CHp_w003}(b).
When $x$ approaches $0.144$,
the primary feature is that new peaks grow rapidly around $q/2\pi\simeq (0.14,0.36)$
and $(0.36,0.14)$ as shown in Fig.~\ref{Fig:CHp_w003}(b).
By checking the FS, see Fig.~3 in Ref.~[\onlinecite{Yuan}],
we conclude that these peaks are induced by the interband excitations,
e.g, from the pocket around $(-\pi,0)$ contributed by the $\alpha$ band to that 
around $(-\pi/2,\pi/2)$ contributed by the $\beta$ band.
From Fig.~\ref{Fig:CHp_w003}(a), we also see that a hump around $q/2\pi=(0.5,0)$
appears for $x=0.144$, which is due to the intra-$\beta$-band excitations, 
e.g., from the FS pocket around $(-\pi/2,\pi/2)$ to the one around $(\pi/2,\pi/2)$.
Note that all the above new peaks are characteristic of the emergence of the
new FS pockets (or the close proximity of the $\beta$ band to the Fermi level).
In addition, the peak seen in the region $0.2<q_x/2\pi<0.3$ in
Fig.~\ref{Fig:CHp_w003}(a), which slightly moves with doping,
comes from the intra-$\alpha$-band excitations between
the old pockets around $(\pi,0)$ and $(-\pi,0)$.

Similarly, we have calculated Im$\chi^{zz}(q,\omega)$ which is qualitatively
the same as Im$\chi_0^{zz}(q,\omega)$ but quantitatively enhanced in most cases.
A few results for the longitudinal susceptibility at a fixed energy
have been shown in Figs.~\ref{Fig:CHp_w003}(c) and \ref{Fig:CHp_w003}(d),
for a comparision with the corresponding transverse one.
We notice that the primary peaks around $q/2\pi\simeq (0.14,0.36)$
and $(0.36,0.14)$ as seen in $\chi^{+-}$ upon increased doping become
much weaker in $\chi^{zz}$, and may even be screened by other peaks nearby.

Finally, we comment on the new peaks around $q/2\pi\simeq (0.14,0.36)$
and equivalent points appearing in the transverse susceptibility. 
As mentioned above, they arise from the interband {\it p-h} excitations.
From the FS plot, we do not see a nesting wave vector connecting two
FS pockets which belong to the two bands, respectively. So the new peaks
found here are not so strong, but in the same order as found in some similar
calculations.\cite{Lu} They should be observable by
neutron scattering which can have measured very weak intensity of
the spin excitations.\cite{Wakimoto}
We point out that all the plots shown in Fig.~\ref{Fig:CHp_w003}
are qualitatively the same for any low energy $\omega/J<0.4$, but the ordinates
are enlarged with increasing $\omega$. Thus the new peaks
may be enhanced by a larger $\omega$, making their observation easier.

In conclusion, we have calculated the dynamical spin susceptibilities
in the AF phase for the electron-doped $t$-$t'$-$t''$-$J$ model. Various results
for the energy and momentum dependences have been given at different dopings.
At low energy, except the collective
spin-wave mode around $(\pi,\pi)$ and $0$, the primary observation 
is that new resonance peaks will appear
around $(0.3\pi,0.7\pi)$ and equivalent points with
increasing doping, which are due to the interband {\it p-h} excitations.
These peaks are pronounced in the transverse susceptibility but not 
in the longitudinal one. We hope that our theoretical results will
stimulate the experimental measurements.

\medskip

We thank F. Yuan for helpful discussions.
This work was supported by the Texas Center for Superconductivity and
Advanced Materials at the University of Houston and the Robert A. Welch Foundation.

\end{document}